\documentclass[prl,nobalancelastpage,twocolumn,superscriptaddress,nolongbibliography]{revtex4-2}

\usepackage{color,amsthm,amsmath,amsxtra,amsfonts,dsfont,graphicx,bm,amssymb}
\usepackage[colorlinks=true,linkcolor=blue, citecolor=blue, urlcolor=blue, bookmarks]{hyperref}
\usepackage{centernot}
\usepackage[dvipsnames]{xcolor}
\usepackage{graphicx}
\usepackage{tikz}
\usepackage{braket}
\usepackage{multirow}
\usepackage{comment}
\usepackage{makecell}

\def \l {\left(} 
\def \r {\right)} 
\def \la {\langle} 
\def \ra {\rangle}  

\newcommand{\be}{\begin{equation}}
	\newcommand{\ee}{\end{equation}}
\newcommand{\bea}{\begin{eqnarray}}
	\newcommand{\eea}{\end{eqnarray}} 
\newcommand{\bse}{\begin{subequations}}
	\newcommand{\ese}{\end{subequations}}

\theoremstyle{plain}

\newtheorem{defn}{Definition}

\newcommand{\prlsection}[1]{{\em {#1}.---~}}

\usepackage{pifont}
\usepackage{tikz}
\usetikzlibrary{decorations.pathreplacing,calligraphy,decorations.markings}

\begin{document}
	\date{\today}

	\newcommand{\bbra}[1]{\<\< #1 \right|\right.}
	\newcommand{\kket}[1]{\left.\left| #1 \>\>}
	\newcommand{\bbrakket}[1]{\< \Braket{#1} \>}
	\newcommand{\pll}{\parallel}
	\newcommand{\nn}{\nonumber}
	\newcommand{\transp}{\text{transp.}}
	\newcommand{\nor}{z_{J,H}}
	
	\newcommand{\hL}{\hat{L}}
	\newcommand{\hR}{\hat{R}}
	\newcommand{\hQ}{\hat{Q}}

\title{Breaking global symmetries with locality-preserving operations}

\begin{abstract}
In the general framework of quantum resource theories, one typically only distinguishes between operations that can or cannot generate the resource of interest. In many-body settings, one can further characterize quantum operations based on underlying geometrical constraints, and a natural question is to understand the power of resource-generating operations that preserve locality. In this work, we address this question within the resource theory of asymmetry,  which has recently found applications in the study of many-body symmetry-breaking and symmetry-restoration phenomena. We consider symmetries corresponding to both abelian and non-abelian compact groups with a homogeneous action on the space of $N$ qubits, focusing on the prototypical examples of $U(1)$ and $SU(2)$.  We study the so-called $G$-asymmetry $\Delta S^{G}_N$, and present two main results. First, we derive a general bound on the asymmetry that can be generated by locality-preserving operations acting on product states. We prove that, in any spatial dimension, $\Delta S^{G}_N\leq (1/2)\Delta S^{G, \rm max}_N[1+o(1)]$, where $\Delta S^{G, \rm max}_N$ is the maximum value of the $G$-asymmetry in the full many-body Hilbert space. Second, we show that locality-preserving operations can generate maximal asymmetry, $\Delta S^{G}_N\sim\Delta S^{G, \rm max}_N$, when applied to symmetric states featuring long-range entanglement. Our results provide a unified perspective on recent studies of asymmetry in many-body physics, highlighting a non-trivial interplay between asymmetry, locality, and entanglement.
\end{abstract}

\author{Michele Mazzoni}
\affiliation{Dipartimento di Fisica e Astronomia, Universit\`a di Bologna and INFN, Sezione di Bologna, via Irnerio 46, 40126 Bologna, Italy}
\author{Luca Capizzi}
\affiliation{Université Paris-Saclay, CNRS, LPTMS, 91405, Orsay, France}
\author{Lorenzo Piroli}
\affiliation{Dipartimento di Fisica e Astronomia, Universit\`a di Bologna and INFN, Sezione di Bologna, via Irnerio 46, 40126 Bologna, Italy}

\maketitle
	

\prlsection{Introduction} In the past two decades, quantum resource theories (QRTs)~\cite{chitambar2019quantum} have significantly influenced our understanding of different quantum phenomena, with entanglement providing the most notable example~\cite{bennett1996concentrating, bennet1996mixed, vedral1997quantifying,horodecki2009quantum}. A QRT is a mathematical framework in which a property of interest (the resource) is indirectly defined in terms of ``free'' states and operations. Generally speaking, free states are those which do not display the property of interest, while free operations are those which do not generate the resource~\cite{chitambar2019quantum}.

In a many-body setting, one can enrich this mathematical framework and ask about the power of resource-generating operations under geometrical constraints. A natural question pertains to operations that preserve locality, as mathematically captured by the notion of quantum cellular automata~\cite{farrelly2019review,arrighi2019overview} and their non-unitary generalizations, the so-called locality preserving (LP) operations~\cite{piroli2020quantum}. LP operations provide very useful toy models for many-body physics, as their repeated application mimics the evolution of many-body systems described by local Hamiltonians and Lindbladians~\cite{farrelly2019review,arrighi2019overview,piroli2020quantum}. At the same time, they can be efficiently implemented on quantum computers~\cite{farrelly2019review}, even in the current era of noisy devices~\cite{preskill2018quantum}, making their study especially timely. 

The power of LP operations is well understood in the resource theory of entanglement. For instance, in analogy with established results on the maximal entanglement growth allowed by local-Hamiltonian dynamics~\cite{calabrese2005evolution,calabrese2006time, osborne2006efficient,eisert2006general,schuch2008entropy,shi2024bounds,toniolo2024dynamical}, it was shown that LP operations can only prepare area-law entangled states~\cite{piroli2020quantum} and are thus limited in the amount of resource they can generate. 

\begin{figure}[h!]
	\centering
    \includegraphics[width = \linewidth]{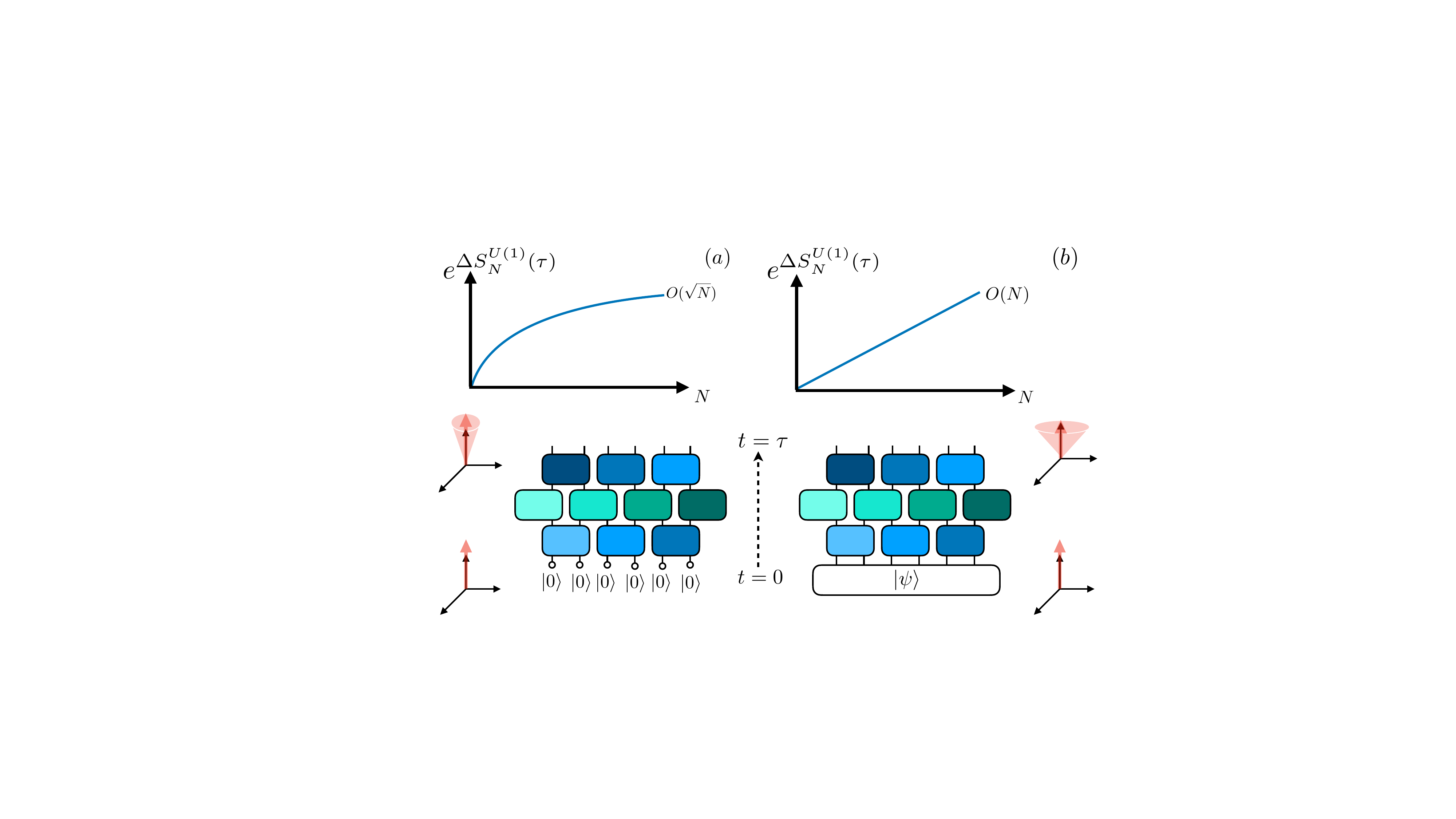}
	\caption{Sketch of our main results. $(a)$ We consider applying LP operations to a product state and bound the asymmetry of the output state  (the cartoon shows a unitary finite-depth circuit in one dimension, but our results hold for generic LP channels and arbitrary dimension). The plot shows the bound on the linearized $G$-asymmetry $e^{\Delta S^{U(1)}_N}=O(\sqrt{N})$. $(b)$ We also show that LP operations can generate maximal asymmetry when applied to symmetric states featuring long-range entanglement. The plot shows the linearized asymmetry $e^{\Delta S^{U(1)}_N}\sim N$ when applying suitably chosen local rotations to an initial Dicke state.}
	\label{fig:results}
\end{figure}

Different from entanglement, the power of LP operations remains largely unexplored in other resource theories and in particular in the resouce theory of asymmetry~\cite{vaccaro2008tradeoff,brs-07,gour2009measuring}. In this QRT, the free operations are the ones that are symmetric with respect to a given symmetry group $G$, while the resources are asymmetric states. This framework is useful when formalizing protocols involving two parties exchanging information that cannot be communicated without some shared reference state~\cite{chitambar2019quantum} (a standard example is the orientation of a spin with respect to a given axis). The two parties can freely perform symmetric operations because these do not require a reference state, while asymmetric states are a resource which can be depleted by (non-unitary) symmetric operations. The asymmetry is quantified by appropriate resource monotones, \emph{i.e.} state functionals that are non-increasing under symmetric operations. 

The asymmetry QRT can be naturally extended beyond communication protocols, and the past few years have witnessed the development of similar ideas in many-body physics~\cite{marvian2014extending,ares2023entanglement} and quantum field theory~\cite{casini2020entropic, casini2021entropic, magan2021, benedetti2024, capizzi2023entanglement}. These developments were motivated by the observation that a particular asymmetry monotone, the $G$-asymmetry $\Delta S^G$~\cite{vaccaro2008tradeoff,gour2009measuring} (also known as entanglement asymmetry~\cite{ares2023entanglement}), can be efficiently computed in interesting physical settings, allowing for several recent studies of symmetry-breaking and symmetry-restoration phenomena~\cite{ares2023lack,rylands2024microscopic, murciano2024entanglement, chalas2024multiple, bertini2024dynamics, rylands2024dynamical, klobas2024asymmetry, caceffo2024entangled,ares2025quantum,foligno2024non, yamashika2024entanglement, yamashika2024quenching, ferro2024non,turkeshi2024quantum, liu2024symmetry,liu2024mpembaloc, banerjee2024asymmetry, klobas2024translation}. An interesting but puzzling observation of some of these studies is the seemingly universal scaling of the $G$-asymmetry in the number of qubits. For example, for a $U(1)$ symmetry acting locally on $N$ qubits, it was found that the asymmetry scales as $\Delta S_N^{U(1)}\sim (1/2)\log N$, namely half of its maximum allowed value, in generic translation-invariant matrix product states~\cite{cv-24}, Gaussian~\cite{ares2023entanglement}, and Haar random states~\cite{ares2024entanglement,ares2025entanglement}.

In this work, we bound the asymmetry generated by LP operations acting on product states, proving that the leading order of $\Delta S^{G}_N$ can be at most half its maximum allowed value, cf. Fig.~\ref{fig:results} for a sketch of our main results. Conversely, we show that LP operations can generate maximal asymmetry when applied to symmetric states featuring long-range entanglement. The key ingredient to derive our bounds is the lightcone of the correlation functions in states prepared by LP operations~\cite{farrelly2019review,piroli2020quantum}, a feature known as the \emph{cluster property}. As the latter is a typical feature of many-body states, our results explain previous observations of the seemingly universal scaling of the $G$-asymmetry and highlight a non-trivial interplay between asymmetry, locality, and entanglement. Our work may also be relevant for experimental studies of asymmetry generation in digital quantum platforms~\cite{joshi2024observing}.

\prlsection{Locality-preserving operations}  We start by briefly reviewing LP operations~\cite{piroli2020quantum}. We consider a regular lattice in $d_L$ dimensions, with $N=M^{ d_L}$ qubits placed at the vertices, and periodic boundary conditions. We define the distance between two qubits, $\delta(n,m)$, as the minimum number of edges connecting the corresponding vertices. The Hilbert space associated with the set of qubits is ${\cal H}=\otimes_{n}{\cal H}_n$, where ${\cal H}_n\simeq \mathbb{C}^2$. Next, given a subset of qubits $A$, we define its $\lambda$-neighborhood, $a_\lambda=\{n : \delta(n,A)\le \lambda\}$, where $\delta(n,A)$ is the minimum distance between the qubit $n$ and any qubit in $A$. We also set  $\bar A_\lambda=A\cup a_\lambda$. Finally, given a subset $C$, we denote by ${\rm Tr}_C$ the trace in $\mathcal{H}_C=\otimes _{j\in C}\mathcal{H}_{j}$ and by $X_C$ an operator supported on $C$. We will also denote by $\ket{0}_j$, $\ket{1}_j$ a basis for the local Hilbert space and by $\sigma^\alpha_j$ the Pauli matrices ($\alpha=x,y,z$). We now introduce LP operations as a special type of quantum channel, \emph{i.e.} trace-preserving completely positive map~\cite{nielsen2010quantum}.
\begin{defn}
${\cal E}_\lambda$ is a LP operation with range $\lambda\in \mathbb{N}$ if, for any sets of qubits $A$ and $B$,
 \be
 \label{eq:defLPQC}
 {\cal E}_{\lambda}^\dagger(X_A Y_{B}) = {\cal E}_{\lambda}^\dagger(X_A) {\cal E}_{\lambda}^\dagger(Y_B)\,,
 \ee
with ${\cal E}_{\lambda}^\dagger(X_A) =X_{\bar{A}_\lambda}$, ${\cal E}_{\lambda}^\dagger(Y_B)=Y_{{\bar B}_\lambda}$.
\end{defn}
\noindent Here, we denoted by ${\cal E}^\dagger$ the adjoint of the channel $\mathcal{E}$, which is defined by the action ${\rm Tr}[{\cal E}^\dagger(A)B]={\rm Tr}[A{\cal E}(B)]$. The simplest LP operations are finite-depth local quantum circuits, \emph{i.e.} unitaries of the form $U^{(D)} =  V_{D} \ldots V_1$ where $V_D$ is a product of two-qubit gates acting on disjoint pairs of neighboring qubits ($D$ is the circuit depth).

In most of this work, we focus on (possibly mixed) initial product states $\rho_0=\otimes_j \omega_j$, where $\omega_j$ is a state in the local Hilbert space $\mathcal{H}_j$. Then, Eq.~\eqref{eq:defLPQC} implies that the output state $\rho=\mathcal{E}_{\lambda}(\rho_0)$ satisfies the cluster property
\be
\label{eq:correlation_clustering}
\la O_i\,O_j\ra_\rho - \la O_i \ra_\rho \la O_j \ra_\rho = 0, \quad \text{for} \quad \delta (i,j) > \Lambda=2\lambda\,,
\ee
where $\braket{O}_\rho={\rm Tr}[\rho O]$, while $O_j$ is an arbitrary operator supported on site $j$. Note that choosing initial product states is very natural both from a theoretical many-body perspective and from the experimental point of view. For instance, current prototypes of quantum computers are typically initialized in unentangled states and evolved by a relatively small number of (noisy) local quantum-circuit steps~\cite{preskill2018quantum,altman2021quantum}, corresponding to LP operations~\cite{farrelly2019review,piroli2020quantum}. 

\prlsection{The $G$-asymmetry} We consider symmetries associated with compact Lie groups and local, homogeneous generators of the form $Q^{\alpha}=\sum_{j}q^\alpha_j$, with $\alpha$ an integer label. We focus on the prototypical examples of $U(1)$ and $SU(2)$, for which we choose an explicit representation: For $U(1)$, we consider the charge $Q=\sum_j (\sigma^z_{j}+1)/2$, while for $SU(2)$ we choose the generators $Q^\alpha=S^{\alpha}=\sum_j\sigma^{\alpha}_j/2$. Now, introducing the $G$-twirling operation
\begin{equation}
\label{eq: G-twirling definition}
\mathcal{G}[\rho] := \int_G \mathrm{d}g\, T(g) \rho\, T(g)^\dagger\,,
\end{equation}
where the integral is with respect to the Haar measure~\cite{mele2024introduction}, the $G$-asymmetry is defined as~\cite{vaccaro2008tradeoff}
\begin{equation}
    \Delta S^{G}_N(\rho)=S_V(\mathcal{G}[\rho] )-S_V(\rho )\,,
\end{equation}
where $S_V(\rho)=-{\rm Tr}(\rho \log \rho)$ is the von Neumann entanglement entropy~\cite{nielsen2010quantum}.

The $G$-asymmetry $\Delta S^{G}_N(\rho)$ is an asymmetry monotone~\cite{schuch2004nonlocal, schuch2004quantum,vaccaro2008tradeoff, gour2009measuring,gour2008resource}. That is, given an arbitrary symmetric channel $\mathcal{E}$, one has $\Delta S^{G}_N(\mathcal{E}[\rho])\leq \Delta S^{G}_N(\rho)$~\footnote{We say that a channel is symmetric if $\mathcal{E}(U_g(\cdot )U_g^\dagger)=U_g\mathcal{E}(\cdot )U_g^\dagger$ for any element $U_{g}$ of the symmetry group}. In fact, the $G$-asymmetry satisfies a stronger condition: given a symmetric channel $\mathcal{E}(\cdot)=\sum_k A_k^\dagger(\cdot) A_k$, with $A_k$ symmetric Kraus operators~\cite{nielsen2010quantum}, one has $\Delta S^{G}_N(\rho)\geq \sum_k p_{k} \Delta S^{G}_N(\tilde \rho_k)$ where $\tilde \rho_k=(A_k^\dagger \rho A_k)/p_k$ and $p_k={\rm Tr}(A_k^\dagger \rho A_k)$. Namely, $\Delta S^{G}_N(\rho)$ does not increase on average under symmetric operations. In addition, it is also easy to show that $\Delta S^{G}_N(\rho)\geq 0$ and $\Delta S^{G}_N(\rho)=0$ if and only if $\rho$ is symmetric, making $\Delta S^{G}_N$ a good measure of asymmetry~\footnote{We mention that in a QRT, one is also typically interested in the connections between a resource monotone and the so-called rates of convertibility between states~\cite{brandao2015reversible, chitambar2019quantum}. In the case of the asymmetry, partial results are known, see \emph{e.g.}, Refs.~\cite{gour2008resource, marvian2014asymmetry,yang2017units,marvian2020coherence,marvian2022operational,yamaguchi2024quantum}.}. 

The maximal value of $\Delta S^{G}_N$ is logarithmic in $N$, with explicit bounds known in the literature~\cite{gour2008resource,gour2009measuring}. For instance, for the $U(1)$ symmetry, one can show~\cite{gour2009measuring}
\begin{equation}\label{eq:bound_u1}
    \Delta S^{U(1)}_N(\rho)\leq \log (N+1)\,.
\end{equation}
Our goal is to understand how the general bounds are modified for states satisfying Eq.~\eqref{eq:correlation_clustering}.

\prlsection{The abelian case} We start by the abelian $U(1)$ symmetry and consider a (generally mixed) state $\rho=\mathcal{E}_{\lambda}(\otimes \omega_j)$, satisfying~\eqref{eq:correlation_clustering}. As a first step, we rewrite
\begin{equation}\label{eq:explicit_form}
    \mathcal{G}[\rho]=\sum_{q=0}^N p_q\tilde{\rho}_q\,,
\end{equation}
where $p_q={\rm Tr}[\rho P_q]$, $\tilde{\rho}_q=P_q\rho P_q/p_q$, while $P_q$ is the projector onto the $q$-eigenspace of the charge $Q$. Eq.~\eqref{eq:explicit_form} follows from the definitions~\eqref{eq: G-twirling definition} and of the Haar measure for $U(1)$~\cite{brs-07,gour2008resource}.

Next, we recall the standard inequality~\cite{nielsen2010quantum}:
\begin{equation}\label{eq:standard_vn_inequality}
S_V\l\sum_a p_a \rho_a\r \le \sum_a p_a S_V(\rho_a) + H(\{p_a\})\,,
\end{equation}
where $\{p_a\}$ is a discrete probability distribution, $\rho_a$ are arbitrary mixed states, while $H(\{p_a\})=-\sum_j p_j\log p_j$ is the Shannon entropy. Applying it to Eq.~\eqref{eq:explicit_form}, we get
\begin{equation}
    \Delta S^{U(1)}_N(\rho)\leq H(\{p_q\})+\sum_q p_q S_V(\tilde{\rho}_q)-S_V(\rho)\,.
\end{equation}
Finally, we observe that $\tilde{\rho}_q$ is obtained from $\rho$ by measuring the charge operator and postselecting on the outcome $q$. On the other hand, the von Neumann entropy does not increase, on average, under measurements~\cite{lindblad1972entropy}, and so $S_V(\rho)\geq \sum_q p_q S_V(\tilde{\rho}_q)$. Putting all together, we have
\begin{equation}\label{eq:bound_shannon_u1}
     \Delta S^{U(1)}_N(\rho)\leq H(\{p_q\})\,.
\end{equation}
Note that this inequality is saturated in the case $\rho$ is pure, and that Eq.~\eqref{eq:bound_shannon_u1} immediately implies the upper bound~\eqref{eq:bound_u1}, because $q$ can take $N+1$ values.

In order to derive a tighter bound for clustering states, a key observation is that Eq.~\eqref{eq:correlation_clustering} constrains the variance of the probability distribution $p_q$, which we denote by $\sigma^2=\mathbb{E}_q[(q-\bar{q})^2]$, with $\bar{q}=\mathbb{E}_q[q]$ and where $\mathbb{E}[\cdot]$ denotes the expectation value. Indeed, using $\braket{Q^2}_\rho=\mathbb{E}_q[q^2]$, $\braket{Q}_\rho=\bar{q}$, and Eq.~\eqref{eq:correlation_clustering} with $O_j=q_j$, we obtain~\cite{SM}
\begin{equation}\label{eq:bound}
    \sigma^2 \le 2 z_\Lambda N\,,
\end{equation}
where $z_\Lambda$ is the cardinality of any of the sets $I_x^\Lambda=\{x^\prime:\delta(x,x^\prime)\leq \Lambda\}$ ($z_\Lambda$ does not depend on $x$). Note that $z_\Lambda$ only depends on $\Lambda$ and the dimensionality of the lattice, but not on $N$.

Eq.~\eqref{eq:bound} allows us to conclude our derivation, using a known bound on the Shannon entropy of discrete random variables. To be precise, let $x$ be an arbitrary integer-valued random variable, with probability function $p_x$ and support $A \subseteq \mathbb{Z}$, such that $0 < \sigma^2 < \infty$. Then, the Shannon entropy is bounded by~\cite{massey1989entropy, rioul2022gaussian} $H(\{p_x\}) < \frac{1}{2}\log\left[2\pi e \left(\sigma^2+\frac{1}{12}\right)\right]$. Combining the above with~\eqref{eq:bound}, we arrive at our first main result
\be\label{eq:first_main_result}
\Delta S^{U(1)}_N (\rho) \le \frac{1}{2}\log\left[2\pi e \left(2z_\Lambda N+\frac{1}{12}\right)\right]\sim \frac{1}{2}\log N\,,
\ee
displaying the announced asymptotic scaling in $N$.

\prlsection{The non-abelian case} The case of the non-abelian $SU(2)$ symmetry turns out to be significantly more involved compared to $U(1)$. Still, it is possible to follow a similar logic, which relies on two main ingredients: ($i$)  the asymmetry is bounded in terms of the Shannon entropy of the probability distribution over the group quantum numbers [cf. Eq.~\eqref{eq:bound_shannon_u1}]; ($ii$); the cluster property~\eqref{eq:correlation_clustering} constrains the latter probability distribution function, resulting in a bound for the corresponding Shannon entropy. 

Following Refs.~\cite{brs-07,gour2009measuring}, we begin by recalling a few facts about the $SU(2)$ representation theory and the $SU(2)$-asymmetry. The eigenvalues of the Casimir operator ${\bf S}^2=(S^{x})^2+(S^{y})^2+(S^{z})^2$ and of $S^z$ are labeled, respectively, by ${\bf S}^2\ket{s,m}=s(s+1)\ket{s,m}$ and $S^z\ket{s,m}=m\ket{s,m}$
For concreteness, we consider the case of $N$ even, so that the allowed values for $s$ and $m$ are $s=0,1,\ldots, N/2$, $m=-N/2, \ldots ,N/2$. We further denote by $P_s$ and $P_m$ the orthogonal projectors onto the subspaces associated with the quantum numbers $s$, $m$.

The integer $s$ labels the $SU(2)$ irreducible representations and $\mathcal{H}$ can be decomposed as
\begin{equation}\label{eq:main_decomposition}
\mathcal{H} = \bigoplus_s \mathcal{M}_s \otimes \mathcal{N}_s\,,    
\end{equation}
where $\mathcal{M}_s$ is the irreducible representation space corresponding to total spin $s$, while $\mathcal{N}_s$ is the associated multiplicity space. Using Schur's lemma, we can write an explicit form for the symmetrized density matrix~\cite{brs-07,gour2009measuring},
\begin{equation}\label{eq:main_tilde_rho_dec}
\mathcal{G}[\rho]=\sum_s p_s \frac{\openone_{\mathcal{M}_s}}{\operatorname{dim} \mathcal{M}_s} \otimes \tilde{\rho}_{s}\,,
\end{equation}
where $p_s={\rm Tr}[\rho P_s]$, while ${\tilde{\rho}}_s=\frac{1}{p_s}{\rm Tr}_{\mathcal{M}_s}[P_s \rho P_s]$.  For pure states, inspection of Eq.~\eqref{eq:main_tilde_rho_dec}, allows one to bound the dimension of the support of $\mathcal{G}[\rho]$, yielding a logarithmic bound~\cite{gour2009measuring} $\Delta S^{SU(2)}_N(\rho)\leq 3\log N (1+o(1))$. We now improve this bound assuming the cluster property.

As a first simplifying observation, we note that given a state $\omega$, we can always find $U=u^{\otimes N}$ such that the state $\rho=U\omega U^\dagger$
satisfies 
\begin{equation}\label{eq:main_vanishing_condition}
\langle  S^x \rangle_{\rho}=\langle  S^y \rangle_{\rho}=0\,.
\end{equation}
This is because $U$ can implement arbitrary rotations of the vector $(\langle S^x\rangle_\omega,\langle S^y\rangle_\omega,\langle S^z\rangle_\omega)$. Importantly, if $\omega$ satisfies the cluster property, so does $\rho$. Then, we can bound the asymmetry of $\rho$, which coincides with that of $\omega$ (since such $U$ is in the image of the $SU(2)$ representation \footnote{Namely, from the right-invariance of the $SU(2)$ Haar measure, $\mathcal{G}[U\omega U^\dagger]=$$\mathcal{G}[\omega]$, cf. Eq. \eqref{eq: G-twirling definition}. Since $S_V(\rho)=S_V(\omega)$, $\Delta S^{SU(2)}_N(\rho)=\Delta S^{SU(2)}_N(\omega)$.}).

Now, starting from~\eqref{eq:main_tilde_rho_dec}, exploiting Eq.~\eqref{eq:standard_vn_inequality}, and the monotonicity of the von Neumann entropy with respect to measurements~\cite{lindblad1972entropy}, we derive~\cite{SM}
\begin{equation}\label{eq:main_ineq_shannon_su2}
\Delta S^{SU(2)}_N(\rho)\leq  \sum_s p_s \log (2s+1)-\sum_{s,m}p_{s,m}\log p_{s,m}
\,.  
\end{equation}
Eq.~\eqref{eq:main_ineq_shannon_su2} generalizes Eq.~\eqref{eq:bound_shannon_u1} to the non-abelian case, and immediately implies $\Delta S^{SU(2)}_N(\rho)\leq 3\log N(1+o(1))$.

As a final step, we bound the RHS of Eq.~\eqref{eq:main_ineq_shannon_su2}, under two constraints induced by the cluster property. The first one is Eq.~\eqref{eq:bound}, which bounds the variance of $Q^z$ and hence of the marginal probability distribution $\{p_m\}$. The second constraint is the following: given a clustering state $\rho$ satisfying~\eqref{eq:main_vanishing_condition}, it is easy to show~\cite{SM}
\begin{equation}\label{eq:second_constraint}
\mathbb{E}[s(s+1)-m^2]=\langle {\bf S}^2 \rangle_\rho-	 \langle (S^{z})^2\rangle_\rho \leq  c(\Lambda) N\,,
\end{equation}
where $c(\Lambda)$ depends on $\Lambda$ but not on $N$. A rigorous bound of the RHS of Eq.~\eqref{eq:main_ineq_shannon_su2} under the constrains~\eqref{eq:bound}, \eqref{eq:main_vanishing_condition}, and \eqref{eq:second_constraint} can be performed using elementary considerations and inequalities, although the derivation is quite long and technical~\cite{SM}. As a result, we obtain
\begin{equation}\label{eq:second_main_result}
    \Delta S_N^{SU(2)}(\rho)\leq \frac{3}{2}\log N+g_\Lambda(N)\,,
\end{equation}
where $g_\Lambda(N)=O(\log \log N))$ is a subleading correction. Eq.~\eqref{eq:second_main_result} states the anticipated $SU(2)$-asymmetry bound for clustering states. In the rest of this letter, we further elaborate on our results and discuss their implications. 

\prlsection{Clustering and typicality} It is important to note that our analysis only requires the cluster property~\eqref{eq:correlation_clustering}, which holds beyond states that are prepared by LP operations. For instance, ground-states of local Hamiltonians satisfy~\eqref{eq:correlation_clustering}, up to exponentially small corrections~\footnote{By inspection, such exponentially small corrections are seen to only give sub-leading corrections to our inequalities, which continue to hold at the leading order in $N$}. As a byproduct, our bounds also apply to those states.

More generally, our work provides a unified perspective on previous studies of asymmetry in many-body systems. For instance, we can revisit the fact that the $U(1)$ asymmetry scales as $\Delta S_N^{U(1)}\sim (1/2)\log N$ in generic translation-invariant matrix product states~\cite{cv-24}, Gaussian~\cite{ares2023entanglement}, and Haar random states~\cite{ares2024entanglement,ares2025entanglement}. Despite being completely different sets of states, they all satisfy the cluster property (up to subleading corrections), so that their asymmetry is bounded by $\sim (1/2)\log N$. On the other hand, in the absence of additional symmetry constraints, it is natural to expect that the bound induced by the cluster property is saturated. We can appreciate the typicality of the $U(1)$-asymmetry  scaling $\sim (1/2)\log N$ for clustering states looking at the probability distribution $p_q$. For typical short-range correlated states, we expect that $p_q$ follows the central limit theorem: heuristically, assuming that the charge densities $q_i$ are i.i.d  variables with variance $\text{Var}[q_i]=\sigma^2$, the total charge $Q=\sum_i q_i$ is distributed as $(Q-\la Q \ra)/\sqrt{N} \sim \mathcal{N}(0,\sigma^2)$, with $\mathcal{N}$ the normal distribution. Such a normal distribution function immediately implies the scaling
$\Delta S^{U(1)}_N\sim (1/2)\log N$~\cite{SM}.

\prlsection{Rotating long-range entangled states}  In light of our results, a very natural question is whether the resource-generation power of LP operations is increased when acting on \emph{entangled} symmetric states. We answer this question in the affirmative: Focusing on the $U(1)$ symmetry, we show that it is possible to obtain the maximal asymmetry scaling $\sim\log N$ when applying LP operations to suitably chosen symmetric entangled states. 

To this end, we first characterize pure states with maximal asymmetry scaling. Note that, by our results, any maximally asymmetric state $\ket{\phi}$ can not be prepared by a finite-depth circuit, and thus must be long-range entangled~\cite{chen2010local,zeng2015quantum}. Going further, given $\ket{\phi}$ corresponding to a probability distribution $\{p_q\}_{q\in A}$, we find that a maximal asymmetry scaling is achieved if $\{p_q\}_{q\in A}$ defines a distribution which is given, at leading order in $N$, by $p_q\sim p(q/N)/N$, where $p(u)\in L^1([0,1])$ is a continuous charge density such that $\int_{[0,1]} du \,p(u)=1$. This condition is better understood in terms of the scaling of the charge generating function $\la e^{i\alpha Q}\ra$, which is the Fourier transform of $p_q$. If $\lim_{N \to \infty}\la e^{i\alpha Q/N}\ra = f(\alpha)$ is a continuous function with a well-defined Fourier transform, then the leading order in the large-$N$ expansion of the charge distribution function is $p(u)= N^{-1}\mathcal{F}[f](u)$ and~\cite{SM}
\be
\Delta S_N^{U(1)} \sim \log N - \int_0^1 du\,p(u)\log p(u).
\ee

It is then easy to show that LP operations can generate maximal asymmetry scaling when applied to entangled symmetric states. As an example, consider the Dicke states~\cite{dicke1954coherence} $\ket{D^z_k}=\binom{N}{k}^{-1/2}(\sum_{j=1}^N \sigma_j^-)^k\ket{0}^{\otimes N}$. Clearly, $\ket{D^z_k}$ has zero $U(1)$ asymmetry. However, when $k$ is extensive, we can map $\ket{D^z_k}$ to a maximally asymmetric state by acting on every qubit with a Hadamard unitary $H$, thus obtaining a Dicke state in the $x$-direction: $\ket{D_k^x}=H^{\otimes N}\ket{D_k^z}$. For $k=N/2$, we have computed the charge density $p(u) = (\pi \sqrt{u(1-u)})^{-1}$, which is smooth and thus yields the announced maximal scaling. We have verified this scaling by an explicit analytic computation, yielding $\Delta S^{U(1)}_N(\ket{D_{N/2}^x}\bra{D^x_{N/2}})\sim \log N + \pi/4$~\cite{SM}.

\prlsection{Outlook}
Our work opens up short-term and long-term research directions. For instance, while we have discussed the case of exponential corrections to the cluster property~\eqref{eq:correlation_clustering}, an immediate question is how our bounds are modified by allowing for power-law violations. It is also very natural to ask whether one can define and characterize a notion of long-range asymmetry, similar to what recently done in the resource-theory of non-stabilizerness~\cite{bravyi2005universal,veitch2014resource, white2021conformal,ellison2021symmetry, korbany2025long}. In the long term, extending the setting of the present work, it would be interesting to systematically study the interplay between locality and resource-generating operations in the general context of quantum-resource theories~\cite{chitambar2019quantum}. We hope that our work will motivate further studies in these directions.

\prlsection{Acknowledgments} We acknowledge discussions and collaboration on related topics with Filiberto Ares, Sara Murciano, and Pasquale Calabrese. The work of MM and LP was funded by the European Union (ERC, QUANTHEM, 101114881). Views and opinions expressed are however those of the author(s) only and do not necessarily reflect those of the European Union or the European Research Council Executive Agency. Neither the European Union nor the granting authority can be held responsible for them.


\let\oldaddcontentsline\addcontentsline
\renewcommand{\addcontentsline}[3]{}
\bibliography{bibliography}
\let\addcontentsline\oldaddcontentsline

\onecolumngrid
\newpage

\appendix
\setcounter{equation}{0}
\setcounter{figure}{0}
\renewcommand{\thetable}{S\arabic{table}}
\renewcommand{\theequation}{S\thesection.\arabic{equation}}

\setcounter{defn}{0}
\setcounter{thm}{0}
\setcounter{figure}{0}

\setcounter{secnumdepth}{2}

\begin{center}
	{\Large \bf Supplemental Material}
\end{center}

Here we provide additional details on the results stated in the main text.

\tableofcontents

\section{Bound on $U(1)$-entanglement asymmetry for clustering states}

In this section we provide additional details on the derivation of the asymmetry bound for clustering states, in the case of the $U(1)$ abelian symmetry. Next, we discuss the example of pure inhomogeneous product states, which saturate the inequality \eqref{eq:bound_shannon_u1}.

\subsection{Derivation of the bound}
As in the main text, we denote the $U(1)$ charge by $Q=\sum_j q_j$, with the charge density $q_j=(\sigma_j^z+1)/2$ supported on $\mathcal{H}_{j}$. We consider a state $\rho$ satisfying the clustering decomposition
\be
\label{eq:SM_correlation_clustering}
\la O_i\,O_j\ra_\rho - \la O_i \ra_\rho \la O_j \ra_\rho = 0, \quad \text{for} \quad \delta (i,j) > \Lambda\,,
\ee
for some integer $\Lambda$, where $\la O\ra_\rho={\rm Tr}[\rho O]$ and $O_j$ is an arbitrary operator supported on $\mathcal{H}_j$. As discussed in the main text, Eq.~\eqref{eq:SM_correlation_clustering} is satisfied by the output of a LP channel acting on a product state.  

We begin by putting a bound on the charge variance. To this end, we first note
\be
\label{eq:SM_bound_connected_2PF}
|\la q_xq_{x'}\ra_c| = |\la q_xq_{x'}\ra-\la q_x\ra \la q_{x'}\ra| \leq |\la q_xq_{x'}\ra|+|\la q_x\ra| \cdot |\la q_{x'}\ra|\leq 2\| q_0\|^2=2\,,
\ee
where $\la q_xq_{x'}\ra_c$ denotes the connected correlation function, while $\| \dots\|$ is the operator norm. Here, we used $| \la \mathcal{O}\ra|\leq \|  \mathcal{O} \|$, $\| \mathcal{O}_1 \mathcal{O}_2\|\leq \| \mathcal{O}_1 \| \cdot \| \mathcal{O}_2\|$, and $||q_j||=1$. Using Eq.~\eqref{eq:SM_correlation_clustering} with $O_j=q_j$ and Eq.\eqref{eq:SM_bound_connected_2PF}, we can bound
\be\label{eq:SM_bound_qsquare}
\la Q^2\ra_c = \left|\sum\nolimits_{x\in \{1,\dots,N\},\,x'\in I_x^\Lambda} \la q_xq_{x'}\ra_c\right| \le \sum_{x\in \{1,\dots,N\},\,x'\in I_x^\Lambda}  |\la q_xq_{x'}\ra_c| \leq 2 z_\Lambda N\,,
\ee
where 
\begin{equation}
    I_x^\Lambda=\{x^\prime:\delta(x,x^\prime)\leq \Lambda\}\,,
\end{equation}
while $z_\Lambda$ is the cardinality of $I_x^\Lambda$ (which is independent of $x$).
That is, the charge variance of a clustering state is at most extensive in the system size. Importantly, the variance of the charge is the second cumulant of the probability distribution $\{p_q\}$, where $p_q = {\rm Tr}[\rho P_q]$ and $P_q$ is the projector onto the $q$-eigenspace of the charge $Q$. Namely,
\begin{align}
\sigma^2 &:= \la Q^2 \ra_c = \sum_q p_q\,(q-\bar{q})^2\,.\label{eq:SM_state_variance}
\end{align}
Therefore, Eqs.~\eqref{eq:SM_bound_qsquare} and ~\eqref{eq:SM_state_variance} immediately yield a bound on the variance of the probability distribution $\{p_q\}$. 

As a next step, we rewrite
\begin{equation}\label{eq:SM_explicit_form}
    \mathcal{G}[\rho]=\sum_{q=0}^N p_q\tilde{\rho}_q\,,
\end{equation}
where $\tilde{\rho}_q=P_q\rho P_q/p_q$, and recall the standard inequality~\cite{nielsen2010quantum}
\begin{equation}\label{eq:von_neuman_h_bound}
S_V\l\sum_a p_a \rho_a\r \le \sum_a p_a S_V(\rho_a) + H(p)\,,
\end{equation}
where $\{p_a\}$ is any discrete probability distribution, $\rho_a$ are arbitrary mixed states, while $S_V(\rho)=-{\rm Tr}[\rho\log\rho]$ and $H(\{p_j\})=-\sum_j p_j\log p_j$ are the von Neumann and Shannon entropies, respectively. Applying this inequality to ~\eqref{eq:SM_explicit_form}, we get
\begin{equation}
    \Delta S^{U(1)}_N(\rho)\leq H(\{p_q\})+\sum_q p_q S_V(\tilde{\rho}_q)-S_V(\rho)\,.
\end{equation}
Now, we observe that $\tilde{\rho}_q$ is obtained from $\rho$ by measuring the charge operator and postselecting on the outcome $q$. On the other hand, the von Neumann entropy does not increase, on average, under measurements~\cite{lindblad1972entropy}, and so $S_V(\rho)\geq \sum_q p_q S_V(\tilde{\rho}_q)$. Putting all together, we have
\begin{equation}\label{eq:SM_bound_shannon_u1}
     \Delta S^{U(1)}_N(\rho)\leq H(q)\,.
\end{equation}

Finally, we make use of a known result in classical probability theory, which is an upper bound on the Shannon entropy of an integer-valued random variable with fixed variance. To be precise, let $x$ be an integer-valued random variable, with probability function $p_x$ and support $A \subseteq \mathbb{Z}$, such that $0 < \sigma^2 < \infty$. Then~\cite{massey1989entropy, rioul2022gaussian}
\be
\label{eq:SM_massey_bound}
H(\{p_x\}) < \frac{1}{2}\log\left[2\pi e \left(\sigma^2+\frac{1}{12}\right)\right]\,.
\ee
Noting that Eq.~\eqref{eq:SM_massey_bound} is a monotonic function of $\sigma$, using Eqs.~\eqref{eq:SM_bound_qsquare} and ~\eqref{eq:SM_state_variance}, we arrive at our final result
\be\label{eq:SM_final_bound}
    \Delta S^{U(1)}_N(\rho)\le \frac{1}{2}\log\left[2\pi e \left(2z_\Lambda N+\frac{1}{12}\right)\right]= \frac{1}{2}\log(N)(1+o(1))\,.
\ee

\subsection{Inhomogeneous product states}

In Ref. \cite{cv-24} it was proven that the asymmetry of translation-invariant product states at large system size $N$ is given by $\Delta S_N^{U(1)} \sim \frac{1}{2}\log N$. In this section, we explicitly obtain the bound \eqref{eq:SM_final_bound} for pure, inhomogeneous product states. The proof we provide is non-trivial, and it relies on the Shepp-Olkin theorem \cite{shepp1981entropy, hillion2017proof}. The computations presented in this section can be considered as an independent check of Eq.~\eqref{eq:SM_final_bound}, but may also be interesting per se.

The state we consider is:
\be
\ket{\{x_i\}} = \bigotimes_{i=1}^N (\sqrt{x_i}\ket{0}_i+ \sqrt{1-x_i}\ket{1}_i), \quad x_i \in [0,1].
\ee
Since the state is not translation-invariant, the local charges $q_i=(\sigma_i^z+1)/2$ are not i.i.d, and the heuristic argument based on the classical central limit theorem cannot be invoked. Nonetheless, the clustering hypothesis hold, as $\la q_i\ra = x_i$, $\la q_i q_j\ra = (1-\delta_{ij})x_ix_j+\delta_{ij}x_i$, implying:
\be
\la q_i q_j \ra = \begin{cases}
x_i - x_i^2, &\quad i=j \\
0, &\quad i\ne j
\end{cases}.
\ee
The charge probabilities $p_q=\bra{\{x_i\}}P_q\ket{\{x_i\}}$ are obtained by taking the Fourier transform of the charge generating function:
\be
\la e^{i\alpha Q}\ra = \prod_{j} (e^{i\alpha} x_j + (1-x_j)).
\ee
The above product is expanded in $2^N$ terms, $\binom{N}{q}$ of which contain a factor $e^{i\alpha q}$ that yields a non-vanishing contribution to $p_q$:
\be
p_q = \int_{-\pi}^{\pi}\frac{d\alpha}{2\pi} e^{-i\alpha q} \prod_{j} (e^{i\alpha} x_j + (1-x_j)) = \underset{1\le j_1 < \dots <j_q \le N}{\sum}\, \underset{\substack{j \in \{j_1,\dots,j_q\}\\ k \notin \{j_1,\dots,j_q\}}}{\prod} x_j(1-x_k).
\ee
We observe that $p_q$ is the probability mass function of a random variable $X=\sum_i X_i$, where $X_i$ are independent Bernoulli variables with parameters $x_i$, $X_i \sim \mathcal{B}(x_i)$. Therefore, since the inequality \eqref{eq:SM_bound_shannon_u1} is saturated, $\Delta S_N^{U(1)}(\ket{\{x_i\}}) = H(X)$, with $H(X)=-\sum_{q=0}^Np_q \log p_q$. The Shepp-Olkin theorem states that the Shannon entropy $H(X)$ is a concave function of the parameters $x_i$, and its unique maximum is reached when all $x_i = 1/2$, yielding $\Delta S_N^{U(1)}(\ket{\{x_i\}})\le \frac{1}{2}\log (N)(1 + o(1))$.

\section{Bound on $SU(2)$-entanglement asymmetry for clustering states}

In this section, we discuss the case of the non-abelian $SU(2)$ symmetry. 

\subsection{Preliminaries and notation}
Following Refs.~\cite{brs-07,gour2009measuring}, we begin by recalling a few notions about the $SU(2)$ representation theory and the $SU(2)$-asymmetry.

As in the main text, we choose the generators $Q^\alpha=S^{\alpha}=\sum_j\sigma^{\alpha}_j/2$, and use the following conventions: the eigenvalues of the Casimir operator ${\bf S}^2=(S^{x})^2+(S^{y})^2+(S^{z})^2$ and of $S^z$ are labeled, respectively by
\begin{align}
	{\bf S}^2\ket{s,m}&=s(s+1)\ket{s,m}\,,\\
	S^z\ket{s,m}&=m\ket{s,m}\,.
\end{align}
For concreteness, we will consider the case of $N$ even. The allowed values for $s$ and $m$ are therefore
\begin{align}
	s&=0,1,\ldots \frac{N}{2}\,,\\
	m&=-\frac{N}{2},\ldots, 0, \ldots \frac{N}{2}\,.
\end{align}
We further denote by $P_s$ and $P_m$ the orthogonal projectors onto the subspaces associated with the quantum numbers $s$ and $m$ (note that $[P_s,P_m]=0$).

The integer $s$ labels the $SU(2)$ irreducible representations and the Hilbert space $\mathcal{H}$ can be decomposed as follows
\begin{equation}\label{eq:decomposition}
\mathcal{H} = \bigoplus_s \mathcal{M}_s \otimes \mathcal{N}_s\,,    
\end{equation}
where $\mathcal{M}_s$ is the representation space corresponding to total spin $s$, while $\mathcal{N}_s$ is the associated multiplicity space.  The dimension of $\mathcal{N}_s$, which corresponds to the multiplicity of the $s$-representation space, can be computed explicitly (see \emph{e.g.}~\cite{Defenu-24}) as
\be
n_s={\rm dim}(\mathcal{N}_s) = \binom{N}{N/2-S}-\binom{N}{N/2-S-1},
\ee
which grows exponentially in $N$.

Using Schur's lemma, we can write an explicit form for the symmetrized density matrix, extending~\eqref{eq:SM_explicit_form} to the case of $SU(2)$. We have in particular~\cite{brs-07,gour2009measuring}
\begin{equation}\label{eq:tilde_rho_dec}
\mathcal{G}[\rho]=\sum_s p_s \frac{\openone_{\mathcal{M}_s}}{\operatorname{dim} \mathcal{M}_s} \otimes \tilde{\rho}_{s}\,,
\end{equation}
where $p_s={\rm Tr}[\rho P_s]$ is the probability distribution over the irreducible representations, while
\begin{equation}
{\tilde{\rho}}_s=\frac{1}{p_s}{\rm Tr}_{\mathcal{M}_s}[P_s \rho P_s]\,.
\end{equation}

By inspection of Eq.~\eqref{eq:tilde_rho_dec}, it is easy to determine the maximum dimension of the support of $\mathcal{G}[\rho]$ and, accordingly, derive the following general logarithmic bound~\cite{gour2009measuring}
\be\label{eq:general_bound}
\Delta S^{SU(2)}_N(\rho) \leq \log\l\sum_s {\rm dim}(\mathcal{M}_s) \times \text{min}\l n_s,{\rm dim}(\mathcal{M}_s)  \r\r\,.
\ee
In addition, Ref.~\cite{gour2009measuring} also exhibited states saturating the bound. Note that ${\rm dim}(\mathcal{M}_s)=2s+1$, so that the RHS of Eq.~\eqref{eq:general_bound} scales as $\sim 3\log N$. In the following subsections, we will prove that, for any state $\rho$ satisfying the clustering decomposition, $\Delta S^{SU(2)}_N(\rho)\leq 3/2\log N + O(\log \log (N))$, namely the asymmetry scaling is bounded by half its maximnum value, as anticipated.

\subsection{Bounding the asymmetry by a classical Shannon entropy}

The decomposition \eqref{eq:tilde_rho_dec} allows us to write 
\be\label{eq:DS_1}
\Delta S^{SU(2)}_N(\rho) = S_V(\mathcal{G}[\rho])-S_V(\rho)=\sum_s p_s \log (2s+1) -\sum_s p_s \log p_s - \sum_s p_s \text{Tr}\l \tilde{\rho}_s \log \tilde{\rho}_s\r-S_V(\rho)\,.
\ee
In this subsection, we bound the right-hand side in terms of a probability distribution function over the quantum numbers $m$ and $s$. 

We first observe that the projector $P_m$ is block-diagonal with respect to the decomposition~\eqref{eq:decomposition}, acting as the identity on the multiplicity space $\mathcal{N}_s$.  In particular, $P_m=\sum_{s} P_{m|s}$ where $P_{m|s} := \tilde{P}_{m|s} \otimes \openone_{\mathcal{N}_s}$ and $\tilde{P}_{m|s}$ is the restriction of $P_m$ onto the subspace $\mathcal{M}_s$. Inserting now appropriate resolutions of the identity, we can rewrite
\begin{align}
\tilde{\rho}_s&=\frac{1}{p_s}
\text{Tr}_{\mathcal{M}_s}\left[\left(\sum_{s'}\sum_{m}P_{m|s'}\right) P_s\rho P_s \left(\sum_{s'}\sum_{m}P_{m|s'} \right) \right]=\frac{1}{p_s}\sum_{m}{\rm Tr}_{\mathcal{M}_s}(P_{m|s} P_s\rho P_s P_{m|s})=\sum_{m}p(m|s)\tilde{\omega}_{s,m}\,;
\end{align}
here, we have defined the conditional probability distribution $p(m|s)=\frac{p_{m,s}}{p_s}$, where $p_{m,s}={\rm Tr}[\rho P_s P_m]$, and the normalized states 
\begin{equation}
    \tilde{\omega}_{s,m}=\frac{{\rm Tr}_{\mathcal{M}_s}(P_{m|s} P_s\rho P_s P_{m|s})}{p_{m,s}}\,.
\end{equation}

Next, we apply the inequality~\eqref{eq:von_neuman_h_bound} to the density matrix $\tilde{\rho}_s$, yielding
\begin{equation}
    S_V(\tilde{\rho}_s) \le \sum_m p(m|s) S_V(\tilde{\omega}_{s,m}) + H(\{p(m|s)\}_m)\,.
\end{equation}
Plugging into~\eqref{eq:DS_1} and using $p(m|s)p_s=p_{m,s}$, we obtain
\begin{align}
\Delta S^{SU(2)}_N(\rho)&\leq \sum_s p_s \log (2s+1) -\sum_s p_s \log p_s +\sum_{s} p_s H(\{p(m|s)\}_m)+\sum_{s,m} p_{s,m}S_{V}(\tilde{\omega}_{s,m})-S_V(\rho)\nonumber\\
&= \sum_s p_s \log (2s+1)-\sum_{s,m}p_{s,m}\log p_{s,m}+\sum_{s,m} p_{s,m}S_{V}(\tilde{\omega}_{s,m})-S_V(\rho)\,.    \label{eq:DS_2}
\end{align}

Now we claim
\begin{equation}\label{eq:intermediate_to_prove}
\sum_{s,m} p_{s,m}S_{V}(\tilde{\omega}_{s,m})-S_V(\rho)\leq 0\,.
\end{equation}
To see this, we first note that $S_{V}(\tilde{\omega}_{s,m})=S_V(\hat{\omega}_{s,m})$, where
\begin{equation}
\label{eq:def_omega_hat}
    \hat{\omega}_{s,m}=\frac{P_{m|s} P_s\rho P_s P_{m|s}}{p_{m,s}}\,.
\end{equation}
Since $\tilde{P}_{m|s}$ is a rank-one projector, $\tilde{P}_{m|s}=|v_m\rangle\langle v_m|$ for some $\ket{v_m}\in \mathcal{M}_s$, then $\hat{\omega}_{s,m}=|v_m\rangle\langle v_m|\otimes \tilde{\omega}_{s,m}$, and $S_V(\hat{\omega}_{s,m})=S_V(\tilde{\omega}_{s,m})$.

 On the other hand, Eq.~\eqref{eq:def_omega_hat} can be rewritten as $\hat{\omega}_{s,m}=\frac{P_{m} P_s\rho P_s P_{m}}{p_{m,s}}$. That is, $\hat{\omega}_{s,m}$ is obtained from $\rho$ by measuring both the Casimir operator ${\bf S}^2=\sum_{\alpha} (S^{\alpha})^2$ and $Q^{z}$, and postselecting on the outcomes $s$ and $m$,  respectively. Then, Eq.~\eqref{eq:intermediate_to_prove} follows from the fact that the von Neumann entropy does not increase, on average, under measurements~\cite{lindblad1972entropy}. 

Putting all together, we arrive at the final result
\begin{equation}  \label{eq:ineq_non_ab2}
\Delta S^{SU(2)}_N(\rho)\leq  \sum_s p_s \log (2s+1)-\sum_{s,m}p_{s,m}\log p_{s,m}
\,.  
\end{equation}
Eq.~\eqref{eq:ineq_non_ab2} generalizes Eq.~\eqref{eq:SM_bound_shannon_u1} to the non-abelian case. It is our starting point to derive the asymmetry bound for states satisfying the cluster property, as detailed in the next subsection.

\subsection{Bound for clustering states}

Given a state $\omega$ satisfying the cluster property with range $\Lambda$, we can always find $U=u^{\otimes N}$ such that the state $\rho=U\omega U^\dagger$
satisfies 
\begin{equation}\label{eq:vanishing_condition}
\langle  S^x \rangle_{\rho}=\langle  S^y \rangle_{\rho}=0\,.
\end{equation}
This is because $U=u^{\otimes N}$ can implement arbitrary rotations of the vector $(\langle S^x\rangle_\omega,\langle S^y\rangle_\omega,\langle S^z\rangle_\omega)$. We will put a bound on the asymmetry of $\rho$, which is the same as that of $\omega$. Importantly,  $\rho$ is also a quantum state satisfying the clustering property with range $\Lambda$. 

For clarity, we will often use the notation $p(s,m)$ in place of $p_{s,m}$. Due to Eq.~\eqref{eq:ineq_non_ab2}, it is enough to put a bound on the functional
\begin{equation}\label{eq:to_bound}
	\mathcal{F}[\{p(s,m)\}]= \sum_{s,m} p(s,m) \log (2s+1) -\sum_{s,m} p(s,m)\log p(s,m)\,.
\end{equation}

As the proof is rather technical, we will split it into a series of intermediate steps.
\begin{itemize}
    \item {\bf Changing variables}.  As a first step, we introduce a new stochastic variable, $\xi$, which is a function of the random variables $s$, $m$. It is defined by 
\begin{equation}
s(s+1)=m^2+\xi^2\,.
\end{equation}
Since $m\leq s$,  we have $s\leq \xi^2\leq s(s+1)$. In principle, $\xi$ could be as large as $\sim N$. However, we will show now that clustering implies an effective bound $\xi\leq O(\sqrt{N})$. In order to make this precise, we first show
\begin{equation}\label{eq:to_prove_1}
\mathbb{E}\left[\xi^2\right]\leq c(\Lambda) N,
\end{equation}
where $c(\Lambda)=O(1)$ is a function of $\Lambda$.

To derive Eq.~\eqref{eq:to_prove_1}, we first observe that the clustering property of $\rho$ implies
\begin{equation}
	\langle {\bf S}^2\rangle_{\rho}-	\sum_{\alpha=x,y,z} \langle S^{\alpha}\rangle_{\rho}^2\leq c(\Lambda) N\,, 
\end{equation}
where $c(\Lambda)=O(1)$ is a function of $\Lambda$ (independent of $N$).  Because of Eq.~\eqref{eq:vanishing_condition}, we get
\begin{equation}
	\langle {\bf S}^2\rangle_{\rho}-	\langle S^{z}\rangle_{\rho}^2\leq c(\Lambda) N\,.
\end{equation}
Now, for any Hermitian operator $A$ and density matrix $\rho$, one has $\langle A^2\rangle_{\rho}\geq \langle A \rangle_\rho^2$, which follows from the Cauchy-Schwarz inequality. Therefore
\begin{align}
	\langle {\bf S}^2 \rangle_\rho-	 \langle (S^{z})^2\rangle_\rho \leq \langle {\bf S}^2\rangle_\rho-	 \langle S^{z}\rangle_\rho^2\leq   c(\Lambda) N\,.
\end{align} 
Finally, we recognize $	\langle {\bf S}^2\rangle_\rho-	 \langle (S^{z})^2\rangle_\rho=\mathbb{E}\left[s(s+1)-m^2\right]=\mathbb{E}\left[
\xi^2\right]$, which concludes the proof of Eq.~\eqref{eq:to_prove_1}.

Now, the values of $(m,s)$ uniquely specify the pair $(m,\xi)$, with $\xi>0$. The converse is also true, as
\begin{equation}\label{eq:s_values}
	s=-\frac{1}{2}+\left(\frac{1}{4}+m^2+\xi^2\right)^{1/2}\,.
\end{equation}
Therefore, we can parametrize $\{p(s,m)\}\rightarrow \{p(m,\xi)\}$, and rewrite~\eqref{eq:to_bound} as
\begin{equation}\label{eq:new_bound}
\mathcal{F}[\{p(m,\xi)\}]=\frac{1}{2} \sum_{m,\xi} p(m,\xi) \log \left[1+4(m^2+\xi^2)\right] -\sum_{m,\xi} p(m,\xi)\log p(m,\xi)\,.
\end{equation}
\item {\bf Restricting the variable range}. Next, we want to use~\eqref{eq:to_prove_1} to induce an effective cutoff on the values of $\xi$ to be considered in the sums appearing in Eq.~\eqref{eq:new_bound}. Choose $\ell>0$ (for the moment, $\ell$ is arbitrary), and rewrite~\eqref{eq:to_prove_1} as [omitting the dependence of $c$ on $\Lambda$]
\begin{equation}
	\sum_{\xi<\ell \sqrt{c N}} p(\xi)\xi^2+ 	\sum_{\xi\geq \ell \sqrt{c N}} p(\xi)\xi^2\leq c N\,.
\end{equation}
Defining $T=	\sum_{\xi\geq \ell \sqrt{c N}} p(\xi)$
it is immediate to show
\begin{equation}
	T (\ell^2 c N)\leq \sum_{\xi\geq \ell \sqrt{c N}} p(\xi)\xi^2\leq c N\,,
\end{equation}
and so
\begin{equation}\label{eq:t_bound}
	T\leq \frac{1}{\ell^2}\,.
\end{equation}
In the following, we denote by $\sum_{m,\xi}'$  and  $\sum_{m,\xi}''$, the sums restricted to $\xi< \ell \sqrt{cN}$ and $\xi\geq \ell\sqrt{cN}$, respectively. We have
\begin{align}
	\frac{1}{2} \sum_{m,\xi} p(m,\xi) \log \left[1+4(m^2+\xi^2)\right]= &	\frac{1}{2} \sideset{}{'}\sum_{m,\xi} p(m,\xi) \log \left[1+4(m^2+\xi^2)\right]+\frac{1}{2} \sideset{}{''}\sum_{m,\xi} p(m,\xi) \log \left[1+4(m^2+\xi^2)\right]\nonumber\\
	\leq 	&\frac{1}{2} \sideset{}{'}\sum_{m,\xi} p(m,\xi) \log \left[1+4(m^2+\xi^2)\right]+\frac{1}{2} T \log [1+N^2+N(N+2)]\nonumber \\
	\leq 	&\frac{1}{2} \sideset{}{'}\sum_{m,\xi} p(m,\xi) \log \left[1+4(m^2+\xi^2)\right]+ T \log (2N)\,.
\end{align}
Similarly, 
\begin{align}
-\sideset{}{''}\sum_{m,\xi} p(m,\xi)\log p(m,\xi)&=-T\sideset{}{''}\sum_{m,\xi}\frac{p(m,\xi)}{T}\log\left[ \frac{p(m,\xi)}{T}T\right]\nonumber\\
&=-T\log T-T\sideset{}{''}\sum_{m,\xi} \frac{p(m,\xi)}{T}\log\left[ \frac{p(m,\xi)}{T}\right]\,.
\end{align}
In the second term, we recognize the entropy for the probability distribution $q(m,\xi)=p(m,\xi)/T$ (it is positive and sums to $1$ by construction). Now, the total number of values that the pair $(m,\xi)$ can take is upper bounded by $(N/2+1)(N+1)\leq N^2$ (assuming $N\geq 4$). Therefore,
\begin{equation}
-	\sideset{}{''}\sum_{m,\xi} \frac{p(m,\xi)}{T}\log\left[ \frac{p(m,\xi)}{T}\right]\leq \log (N^2)\,.
\end{equation}
Putting all together
\begin{align}
	-\sum_{m,\xi} p(m,\xi)\log p(m,\xi)&=	-\sideset{}{'}\sum_{m,\xi} p(m,\xi)\log p(m,\xi)	-\sideset{}{''}\sum_{m,\xi} p(m,\xi)\log p(m,\xi)\nonumber\\
	\leq& 	-\sideset{}{'}\sum_{m,\xi} p(m,\xi)\log p(m,\xi) -T\log T +T\log (N^2)\,.
\end{align}

Collecting all terms,
\begin{align}
	\mathcal{F}[\{p(m,\xi)\}]&\leq -T\log T  + T\log (2N^3) \nonumber\\
	+& \frac{1}{2} \sideset{}{'}\sum_{m,\xi}  p(m,\xi) \log \left[1+4(m^2+\xi^2)\right] -\sideset{}{'}\sum_{m,\xi}  p(m,\xi)\log p(m,\xi)\,,
\end{align}
and exploiting~\eqref{eq:t_bound}, we arrive at
\begin{align}\label{eq:final_form_bound}
	\mathcal{F}[\{p(m,\xi)\}]&\leq \frac{1}{2}\log 2  + (\ell^{-2})\log (2N^3) \nonumber\\
	+& \frac{1}{2} \sideset{}{'}\sum_{m,\xi}  p(m,\xi) \log \left[1+4(m^2+\xi^2)\right] -\sideset{}{'}\sum_{m,\xi}  p(m,\xi)\log p(m,\xi)\,,
\end{align}
where we also used $-T\log T \leq (1/2)\log 2$ for $0\leq T\leq 1$.
\item {\bf Bounding the restricted sums}. As a last step, we put a bound on the terms appearing in the second line of Eq.~\eqref{eq:final_form_bound}. Although this might appear as hard as bounding the functional~\eqref{eq:to_bound}, the task is now greatly simplified because, by construction, the random variable $\xi$ takes value in a relatively small set.

We now take the most delicate step of the proof. We first rewrite
\begin{align}
 -&\sideset{}{'}\sum_{m,\xi}  p(m,\xi)\log p(m,\xi)= -\sum_{m}  \sum_{s\in \mathcal{S}_{\ell,m}}p(s,m)\log p(s,m)\nonumber\\
 =&-\sum_{m}  p(m)\sum_{s\in \mathcal{S}_{\ell,m}}\frac{p(s,m)}{p(m)}\log\left[ \frac{p(s,m)}{p(m)}\right]-\sum_m \sum_{s\in \mathcal{S}_{\ell,m}}p(s,m)\log p(m)\nonumber\\
\leq &-\sum_{m}  p(m)\log p(m)-\sum_{m}  p(m)\sum_{s\in \mathcal{S}_{\ell,m}}\frac{p(s,m)}{p(m)}\log\left[ \frac{p(s,m)}{p(m)}\right]\,.
 \label{eq:temp_eq}
\end{align}
Here, we define $\mathcal{S}_{\ell,m}=\{s: \xi(s,m)< \ell \sqrt{c(\Lambda)N}\}$, and we used
\begin{equation}
\sum_{m}  p(m)\log p(m)-\sum_m \sum_{s\in \mathcal{S}_{\ell,m}}p(s,m)\log p(m)\leq 0\,.
\end{equation}
Next, define $q_m(s)=\frac{p(s,m)}{p(m)}$. Clearly, $0\leq q_m(s)\leq 1$ and $\sum_{s\in \mathcal{S}_{\ell,m}}q_m(s)\leq 1$. We can then apply the following fact, whose proof is immediate: for any $\{q_k\}_{k=1}^R$ such that $0\leq q_k\leq 1$ and $\sum_{k}q_k\leq 1$, it holds that $-\sum_{k=1}^R  q_k\log q_k\leq \log (R+1)$.
In order to apply this result to Eq.~\eqref{eq:temp_eq}, we need to bound the number of $s$-values  that $q_m(s)$ is supported on, for a given $m$. This can be done precisely because we have a cutoff on the allowed values of $\xi$.

Explicitly, we start from Eq.~\eqref{eq:s_values} and rewrite it as
\begin{equation}\label{eq:new_s_values}
	s=-\frac{1}{2}+\left(m^2+\frac{1}{4}\right)^{1/2}+\left\{\left(\frac{1}{4}+m^2+\xi^2\right)^{1/2}-\left(m^2+\frac{1}{4}\right)^{1/2}\right\}\,.
\end{equation}
From this expression, it is clear that, for a fixed $m$, the number of values that $s$ can take is bounded by the maximum value that
\begin{equation}
	\Delta(\xi)= \left\{\left(\frac{1}{4}+m^2+\xi^2\right)^{1/2}-\left(m^2+\frac{1}{4}\right)^{1/2}\right\}\,,
\end{equation}
can take, as a function of $\xi$. This is because 	$\Delta(\xi)\geq 0$ and $\Delta(\xi)$ can only take discrete values, separated by $1$. Defining $a=\left(m^2+1/4\right)^{1/2}$, we get
\begin{equation}
	\Delta(\xi)=  \left\{\left(a^2+\xi^2\right)^{1/2}-|a|\right\}=|a|\left\{(1+\xi^2/a^2)^{1/2}-1\right\}\,,
\end{equation}
and using $\sqrt{1+x^2}\leq 1+(x^2/(1+x))$ for $x>0$,
\begin{equation}
	\Delta(\xi)\leq   |a|\left\{\left(1+\frac{\xi^2/a^2}{1+(\xi/a)}\right)-1\right\}=\frac{\xi^2}{\xi+\left(m^2+1/4\right)^{1/2}}\,.
\end{equation}
Recalling $\xi\leq \ell \sqrt{cN}$, we arrive at
\begin{equation}
		\Delta(\xi)\leq \frac{\ell ^2 cN }{\ell \sqrt{cN}+\left(m^2+1/4\right)^{1/2}}\,.
\end{equation}

Finally, putting all together, Eq.~\eqref{eq:final_form_bound} yields
\begin{align}
 -&\sum_{m}  \sum_{s\in \mathcal{S}_{\ell,m}}p(s,m)\log p(s,m)\nonumber\\
	\leq &-\sum_{m}  p(m)\log p(m)+\sum_{m}  p(m)\log \left(1+\frac{\ell ^2 cN }{\ell \sqrt{cN}+\left(m^2+1/4\right)^{1/2}}\right)\,.
\end{align}

Similarly, the term of the second line in Eq.~\eqref{eq:final_form_bound} can be bounded by
\begin{equation}
\frac{1}{2} \sideset{}{'}\sum_{m,\xi}  p(m,\xi) \log \left[1+4(m^2+\xi^2)\right]\leq \frac{1}{2}\sum_m p(m)\log[1+4(m^2 +\ell^2 c N)]\,,
\end{equation}
where we used $\sideset{}{'}\sum_{\xi}  p(m,\xi) \leq p(m)$.

\item {\bf The final result}. 
In summary, we arrived at the following result: for any $\ell>0$, we have
\begin{align}\label{eq:final_form}
		\mathcal{F}[\{p(m,\xi)\}]&\leq \frac{1}{2}\log 2 + (\ell^{-2})\log (2N^3)-\sum_{m}  p(m)\log p(m) +\sum_{m}p(m) f_\ell(m)\,,
\end{align}
where
\begin{equation}
		f_\ell(m)=\frac{1}{2}\log[1+4(m^2 +\ell^2 c N)]+\log \left(1+\frac{\ell ^2 cN }{\ell \sqrt{cN}+\left(m^2+1/4\right)^{1/2}}\right)\,.
\end{equation}

So far, we have not specified $\ell$. Let us now fix $\ell=\log(N)$. With this choice, the second term in the RHS. of Eq.~\eqref{eq:final_form} is trivial. Indeed $(\ell^{-2})\log (2N^3)=O([\log N]^{-1})<1$, for sufficiently large $N$. Next, the equation $\partial f_{\log N}(m)/\partial m=0$ has three real solutions, $m=0$ (local maximum) and $m=\pm m_*$ (local minima). The analytic expression of $m_*=m_*(N,c)$ is quite cumbersome and we do not report it here. Thus, it holds:
\be
f_{\log N}(m) \le \max\{f_{\log N}(0),\, f_{\log N}(N/2)\}.
\ee
We see that both $f_{\log N}(0)=\log N + O(\log \log N)$, $f_{\log N}(  N/2)=\log N + O(\log \log N)$, with $|f_{\log N}(0)-f_{\log N}(  N/2)| = O(\log\log N)$.
It follows that $	f_{\log N}(m)\leq \log N + O(\log \log N)$ and so
\begin{equation}
\sum_{m}p(m) f_\ell(m)\leq \log N + O(\log \log N)\,.
\end{equation}
Finally, when studying the $U(1)$ case, we have already shown that the cluster property of $\rho$ implies $-\sum_{m}  p(m)\log p(m)\leq (1/2) \log N +d$, where $d=O(1)$ depends on $\Lambda$. 
 
 Putting all together, we have therefore proven
 \begin{equation}
 			\mathcal{F}[\{p(m,\xi)\}]\leq \frac{3}{2}\log N +O[\log \log (N)]\,,
 \end{equation}
which is the anticipated result.

\end{itemize}

\section{Maximally asymmetric states}

In this section we provide additional details on pure states displaying maximal asymmetry, focusing on the case of $U(1)$.

At any finite $N$, the maximal $U(1)$-asymmetry is obtained when the charge distribution $p_q$ is flat~\cite{gour2008resource}, yielding $\Delta S_N^{U(1)}=\log (N+1)$. We now show that the same leading scaling of $\Delta S$ can be obtained whenever the probability distribution of the charge density $Q/N$ is continuous in the thermodynamic limit. It is important to stress that such a scenario is not typical in the context of gapped ground states or thermal states (in the absence of spontaneous symmetry breaking). For the aforementioned cases, clustering of correlations holds, implying the existence of a scaled cumulant generating function in the limit $N \rightarrow \infty$
\be
\label{eq:SM scaled cumulant generating function}
\frac{1}{N}\log\la e^{i\alpha Q}\ra = \sum^{\infty}_{n=0} \frac{(i\alpha)^n}{n!}\la Q^n\ra_c /N,
\ee
with $\la Q^n\ra_c$ the $n$-th cumulant of $Q$. In other words, the cumulants of $Q$ are extensive, and $(Q-\la Q \ra)/\sqrt{N}$ is normal-distributed in the large $N$ limit, with a variance denoted by $\sigma^2$. The generating function can be expanded at leading order as
\be
\label{eq:SM characteristc function gaussian}
\la e^{i\alpha Q}\ra \simeq e^{i\alpha \la Q\ra} e^{-\frac{N \alpha^2 \sigma^2}{2}}
\ee
and the probability distribution is
\begin{align}
p_q = \bra{\psi}P_q \ket{\psi} &= \int_{-\pi}^\pi \frac{d\alpha}{2\pi} e^{-i\alpha q}\la e^{i\alpha Q}\ra \nonumber \\&\underset{N \text{\,large}}{\simeq} \int_{-\infty}^{\infty} \frac{d\alpha}{2\pi} e^{i\alpha (q-\la Q \ra)}e^{-\frac{N\alpha^2 \sigma^2}{2}} = \frac{1}{\sqrt{2\pi N \sigma^2}} \exp{\left[-\frac{(q-\la Q\ra)^2)}{2 N \sigma^2}\right]}.
\end{align}
As a consequence, one can show that the asymmetry of a pure state, being the Shannon entropy of the probability distribution, satisfies $\Delta S = H(\{p_q\}) \simeq 1/2 \log N$.

We now discuss a general mechanism that leads to the saturation of the bound \eqref{eq:SM_final_bound} at the leading order. Let us consider the generating function of the charge density in the large $N$ limit
\be
\label{eq:SM scaling char function maximal asymmetry}
\underset{N\rightarrow \infty}{\lim}\left\langle e^{i\omega \frac{Q}{N}}\right\rangle =  f(\omega).
\ee
We assume that the Fourier transform of $f(\omega)$, which gives the probability distribution of the charge density, exists and it is continuous: we stress that this is not the case when central limit theorem applies, since the charge density is $\delta$-distributed (in the $N\rightarrow\infty$ limit).

Given $p(u)$ the Fourier transform of $f(\omega)$, we can approximate $p_q \simeq p(q/N)/N$ in the large $N$ limit and estimate the asymmetry as
\be
\Delta S^{U(1)}_N = -\sum_{q=0}^{N} p_q \log p_q \simeq \log N - \int_{0}^1 du\  p(u) \log p(u),
\ee
where we used the fact that $\int_{0}^1du\,p(u)=\sum_q p_q =1$. It is worth noting that, in many interesting cases, $p(u)$ might show edge singularities: as a consequence $\int_0^1 du \ p(u)^n$ is not necessarily finite (that is, $p(u)\in L^1([0,1])$ but in general $p(u) \notin L^n([0,1])$ for $n>1$). In those cases, subleading logarithmic contributions can arise in the computation of the $n$th R\'enyi asymmetry, for $n>1$.

Here, we analyze two paradigmatic examples: the ``kink'' and Dicke states. Kink states have a flat probability distribution of the charge, and thus they have maximal asymmetry. Dicke states, on the other hand, are symmetric, but because of their long-range correlations, we show how it is possible to achieve maximal asymmetry $\Delta S^{U(1)}_N \simeq \log N$ by applying local rotations.

\subsubsection{Kink states}
We define a kink state as
\be
\ket{K} :=\frac{1}{\sqrt{N}} \sum_{j=1}^N \ket{j}, \quad \ket{j}= \ket{0\dots 0\underset{j}{1}\dots 1},
\ee
which is a uniform superposition of states with charges $j=1,\dots, N$. The charge probability distribution is uniform:
\be
p_q = \bra{K}P_q \ket{K} = \frac{1}{N}, \quad \forall \, q=1,\dots, N.
\ee
Hence, kink states have maximal asymmetry $\Delta S^{U(1)}_N = \log (N+1)$ at any finite size $N$. To make contact with the mechanism explained previously, we compute the continuous distribution $p(u)$ from the generating function of $Q$:
\be
\bra{K}e^{i\alpha Q}\ket{K} = \frac{1}{N}\sum^{N}_{j=1} e^{i\alpha (j-1)} = \frac{e^{i\alpha N}-1}{N(e^{i\alpha}-1)}.
\ee
In the thermodynamic limit $N \rightarrow \infty$ with $\omega=\alpha N$ fixed, we obtain
\be
\bra{K}e^{i\alpha Q}\ket{K} =\frac{e^{i\omega}-1}{i\omega} + O\l\frac{1}{N}\r.
\ee
We then Fourier transform the function above (at leading order), obtaining a flat probability distribution of the charge density $u$
\be
p(u) = \int^{\infty}_{-\infty} \frac{d\omega}{2\pi} e^{-i\omega u} \cdot \frac{e^{i\omega}-1}{i\omega} = \chi_{[0,1]}(u).
\ee

\subsubsection{Rotated Dicke states}

Finally, we show how Dicke states, which are entangled states with definite spin in the $z$-direction and thus have zero asymmetry with respect to the charge $Q^z$, can be mapped into maximally asymmetric states by acting on them with a one-layer circuit consisting of local unitaries. Namely, it is only required to rotate the state into a Dicke state in the $x$-direction, which -at large system size- has maximal asymmetry with respect to spin measurements in the $z$-direction. We provide two complementary proofs of this: first, we explicitly derive the discrete charge distribution at finite size and compute the large-volume limit of the asymmetry. Then, we show how the limiting distribution of the charge density in the large-volume limit is obtained from the generating function of the $U(1)$ charge.

Dicke states in the $z$-direction on an $N$-sites lattice are defined in the following way:
\be
\label{eq:SM Dicke states z direction def 1}
\ket{D_k^z} = \binom{N}{k}^{-1/2} \underset{1\le j_1 < \dots < j_k \le N}{\sum} \sigma_{j_1}^x\cdots \sigma_{j_k}^x \ket{0}^{\otimes N}, \quad k=0,\dots, N,
\ee
that is, $\ket{D_k^z}$ is the normalized sum of all the distinct product states made with $k$ qubits $\ket{1}$ and $N-k$ qubits $\ket{0}$. A Dicke state in the $x$-direction, denoted by $\ket{D^x_k}$, is obtained by rotating each qubit with a Hadamard matrix:
\be
H= \frac{1}{\sqrt{2}}\begin{pmatrix} 1 & 1 \\ 1  &-1\end{pmatrix}, \quad H\ket{0} =\ket{+}, \quad H\ket{1} = \ket{-},
\ee
where $\ket{+}$, $\ket{-}$ are the eigenstates of $\sigma^x$:
\be
\ket{\pm} = \frac{\ket{0}\pm \ket{1}}{\sqrt{2}}, \quad \sigma^x\ket{\pm} = \pm\ket{\pm}.
\ee

In \cite{bernard2024dynamical} it was shown that each Dicke state in the $x$-direction is a linear combination of Dicke states in the $z$-direction:
\begin{equation}
\ket{D_k^x} = H^{\otimes N}\ket{D_k^z} = \frac{1}{2^{N/2}}\sum_{i=0}^N \sqrt{\binom{N}{i}\binom{N}{k}}K_i\left(k;\frac{1}{2},N\right)\ket{D_i^z},
\end{equation}
where $K_i\left(k;\frac{1}{2},N\right)$ are symmetric Krawtchouk polynomials
\begin{equation}
K_i\left(k;\frac{1}{2},N\right) := {}_{2}F_1(-i,-k,-N;2) = \binom{N}{i}^{-1}\sum_{j=0}^i(-1)^n \binom{N-k}{i-j}\binom{k}{j}, 
\end{equation}
that satisfy the orthogonality relation \footnote{The normalization we adopted for the Krawtchouk polynomials is different from the one in \cite{bernard2024dynamical}, and it is chosen so to ensure that $\braket{D_k^x|D_l^x} = \delta_{k,l}$.}:
\begin{equation}
\label{eq:Krawtchouk_orthogonality}
\sum_{i=0}^N \binom{N}{i} K_i\left(k;\frac{1}{2},N\right) K_i\left(l;\frac{1}{2},N\right) = 2^N \delta_{k,l} \binom{N}{k}^{-1}.
\end{equation}
 As a simple check of the above transformation law, one can for instance verify that (here $N=3$)
\begin{equation*}
\ket{D_1^x} = \sqrt{\frac{3}{8}} \ket{D_0^z} + \sqrt{\frac{1}{8}} \ket{D_1^z} - \sqrt{\frac{1}{8}} \ket{D_2^z} - \sqrt{\frac{3}{8}} \ket{D_3^z},
\end{equation*}
as obtained from a direct expansion of the states $\ket{++-},\, \ket{+-+},\, \ket{-++}$ in the $\{\ket{0}, \ket{1}\}$ basis. 

We are interested in the probability $p_k(q;N)$ of obtaining a value $q=0,\dots,N$ when performing a measurement of the charge $Q^z$ in the rotated Dicke state $\ket{D_k^x}$. Let us denote as usual by $P_q$ the projector in the subspace of fixed charge $q$. Then $P_q \ket{D_i^z} = \delta_{q,i} \ket{D_i^z}$, and the probability $p_k(q;N)$ is given by:
\begin{align}
p_k(q;N) &= \braket{D_k^x|P_q|D_k^x}= \frac{1}{2^N} \binom{N}{k}\sum_{i,j=0}^N \sqrt{\binom{N}{i}\binom{N}{j}}  K_i\left(k;\frac{1}{2},N\right) K_j\left(k;\frac{1}{2},N\right) \braket{D_i^z|P_q|D_j^z} \nonumber \\
&=\frac{1}{2^N} \binom{N}{k}\binom{N}{q} K_q\left(k;\frac{1}{2},N\right) K_q\left(k;\frac{1}{2},N\right),
\end{align}
where \eqref{eq:Krawtchouk_orthogonality} implies:
\begin{equation}
\sum_{q=0}^N p_k(q;N) = 1.
\end{equation}

Obtaining the asymptotic scaling of the asymmetry:
\begin{equation}
\Delta S^{U(1)}_N = -\sum_{q=0}^N p_k(q;N) \log p_k(q;N),
\end{equation}
at large $N$ and for arbitrary values of $k$ is a hard task to perform analytically, due to the fast oscillatory behavior of Krawtchouk polynomials, which results in asymptotic expansions of $K_q\left(k;\frac{1}{2},N\right)$ strongly dependent on the scaling of $k$ with $N$ \cite{li2000uniform, dominici2008asymptotic}. However, we observe numerically that, when $k$ is fixed in the large-volume limit, $\Delta S \simeq \frac{1}{2}\log N$, whereas if the ratio $k/N$ is fixed $\Delta S \simeq \log N$. Some simplifications occur in the case $k=N/2$, making the large-volume limit analytically tractable. Indeed, if $N=2M$, $k=M$:
\begin{equation}
\label{eq:SM Dicke state probabilities discrete complete}
p_M(q;2M) = \frac{1}{2^{2M}}\binom{2M}{M}\binom{2M}{q}^{-1}\left[\sum_{j=0}^q(-1)^j \binom{M}{q-j}\binom{M}{j}\right]^2 = \begin{cases}
0\quad &q \,\text{odd}\\
\frac{1}{2^{2M}}\binom{2M}{M}\binom{2M}{q}^{-1}\binom{M}{q/2}^2 \quad &q \,\text{even}
\end{cases}.
\end{equation}
Thus, in this case, the asymmetry reduces to
\begin{equation}
\Delta S^{U(1)}_N= -\sum_{q=0}^M p_M(2q;2M) \log p_M(2q;2M).
\end{equation}
By fixing $q = uM$, $u=0,1/M,2/M,\dots,1$, the Stirling approximation yields the large-$N$ leading behavior:
\begin{equation}
\label{eq:SM Dicke state charge density}
p_M(2uM;2M) \simeq \frac{1}{\pi M \sqrt{u(1-u)}},
\end{equation}
which implies
\be
\Delta S^{U(1)}_N \simeq - M\int_0^1 \mathrm{d}u\, p_M(2uM;2M)\log p_M(2uM;2M) 
= \frac{1}{\pi} \int_0^1 \mathrm{d}u\, \frac{\log (\pi M \sqrt{u(1-u)})}{\sqrt{u(1-u)}} = \log N + \log \frac{\pi}{4}.
\ee
The integral is convergent and it is easily performed via the change of variable $u =\sin^2\theta$. We numerically verified that the result above reproduces the entanglement asymmetry with an error $O(10^{-4})$ at $N=1000$.

We now follow a complementary approach and obtain the charge density $p(u)$ in the limit of large $N$ by computing the Fourier transform of the charge generating function \footnote{Notice that we are now swapping the roles of the generators $Q^x$ and $Q^z$. This is just for convenience, and it does not affect the result on the charge distribution in the rotated Dicke state.}:
\be
\la e^{i\alpha Q^x} \ra := \bra{D^z_k}e^{i\alpha Q^x} \ket{D^z_k}, \quad Q^x = \sum_j \frac{\sigma^x_j + 1}{2}.
\ee
In order to do so, we observe that the Dicke state in \eqref{eq:SM Dicke states z direction def 1} can be written as:
\be
\ket{D_k^z} = \binom{N}{k}^{-1/2} (\sigma^-)^k \ket{0}^{\otimes N}, \quad \sigma^{\pm} = \sum_j \sigma^{\pm}_j := \sum_j \frac{\sigma_j^x \pm i\sigma_j^y}{2}.
\ee
As before, we are interested in the limit where $k/N$ is kept fixed in the thermodynamic limit, as it is in this regime that the state $\ket{D_k^z}$ displays long-range correlations.

We compute the generating function of $Q^x$ as 
\be\label{eq:g_fun_scar}
\la e^{i\alpha Q^x}\ra \propto \oint \frac{d\bar{\zeta}'}{2\pi i \bar{\zeta}'} \oint \frac{d\zeta}{2\pi i \zeta} \zeta^{-k} \bar{\zeta'}^{-k} \bra{\Uparrow} \exp(\bar{\zeta}'\sigma^+) \exp(i\alpha Q)\exp(\zeta \sigma^-)\ket{\Uparrow},
\ee
where $\ket{\Uparrow}:= \ket{0}^{\otimes L}$ and one can easily check that the proportionality constant in the right-hand side is $N!\,k!\,(N-k)!$. In addition, a simple calculation shows
\be\label{eq:overlap}
\bra{\Uparrow} \exp(\bar{\zeta}'\sigma^+) \exp(i\alpha Q^x)\exp(\zeta \sigma^-)\ket{\Uparrow} = \l \frac{e^{i\alpha}+1}{2}(1+\bar{\zeta'}\zeta) + \frac{e^{i\alpha}-1}{2}(\bar{\zeta'}+\zeta)\r^N.
\ee
Since the right-hand side of the above equation is a polynomial in $\zeta,\bar{\zeta'}$, the computation of \eqref{eq:g_fun_scar} boils down to the identification of the coefficient of $\zeta^{k} \bar{\zeta'}^{k}$. Similarly, $\bra{D^z_k}e^{i\alpha Q^x}\ket{D^k_z}$ is a polynomial in $e^{i\alpha}$ and $p_k(q;N)$ is identified by the coefficient of $e^{i\alpha q}$.

We calculate \eqref{eq:g_fun_scar} in the thermodynamic limit $N\rightarrow \infty$, with $\alpha N$ fixed. In this limit, one employs saddle-point techniques (as in Ref. \cite{mcfm-25}): in particular, for $k/N = 1/2$ the integral localizes at $\zeta' = \zeta$ and $|\zeta|=1$, yielding
\be
\bra{D^z_k}e^{i\alpha Q^x}\ket{D^k_z} \simeq \oint_{|\zeta|=1}\frac{d\zeta}{2\pi i \zeta} e^{i\alpha N (1-\text{Re}(\zeta))/2} = \int^{2\pi}_0\frac{d\theta}{2\pi}e^{i\alpha N (1-\cos \theta)/2} = e^{i\alpha N/2} J_0(\alpha N/2).
\ee
Thus, the probability distribution of the charge density, obtained as the Fourier transform of the generating function, is
\be
\label{eq:SM Dicke state charge density scaled}
p(u) = \int^{\infty}_{-\infty} \frac{d\omega}{2\pi} \int^{2\pi}_0 \frac{d\theta}{2\pi} e^{i\omega(u(\theta)-u)} = \int^{2\pi}_0 \frac{d\theta}{2\pi} \delta(u-u(\theta))= \frac{1}{\pi \sqrt{u(1-u)}},
\ee
where $u(\theta) = (1-\cos\theta)/2$. As expected, $p(u)/M=p_M(2uM;2M)$, with $M=N/2$ and $p_M(2uM;2M)$ given in \eqref{eq:SM Dicke state charge density}.

\end{document}